\documentclass[aps,prd,twocolumn,superscriptaddress,amsmath,nofootinbib]{revtex4-1}

\usepackage{aas_macros}
\usepackage{graphicx}
\usepackage[hidelinks]{hyperref}

\usepackage{cleveref}
\Crefname{equation}{Equation}{Equations}
\crefname{equation}{Eq.}{Eqs.}
\Crefname{figure}{Figure}{Figures}
\crefname{figure}{Fig.}{Figs.}
\crefname{table}{Tab.}{Tabs.}
\Crefname{table}{Table}{Tables}
\crefname{section}{Sec.}{Secs.}
\Crefname{section}{Section}{Sections}

\usepackage{color}

\begin{document}

\title{Rapid numerical solutions for the Mukhanov-Sasaki equation}

\author{W.~I.~J.~Haddadin}
\email[]{wh293@cam.ac.uk}
\affiliation{Kavli Institute for Cosmology, Madingley Road, Cambridge, CB3 0HA, UK}
\affiliation{Astrophysics Group, Cavendish Laboratory, J.J.Thomson Avenue, Cambridge, CB3 0HE, UK}
\affiliation{King's College, King's Parade, Cambridge CB2 1ST, UK}

\author{W.~J.~Handley}
\email[]{wh260@mrao.cam.ac.uk}
\affiliation{Kavli Institute for Cosmology, Madingley Road, Cambridge, CB3 0HA, UK}
\affiliation{Astrophysics Group, Cavendish Laboratory, J.J.Thomson Avenue, Cambridge, CB3 0HE, UK}
\affiliation{Gonville \& Caius College, Trinity Street, Cambridge, CB2 1TA, UK}

\begin{abstract}
    We develop a novel technique for numerically computing the primordial power spectra of comoving curvature perturbations. By finding suitable analytic approximations for different regions of the mode equations and stitching them together, we reduce the solution of a differential equation to repeated matrix multiplication. This results in a wavenumber-dependent increase in speed which is orders of magnitude faster than traditional approaches at intermediate and large wavenumbers. We demonstrate the method's efficacy on the challenging case of a stepped quadratic potential with kinetic dominance. We further generalise to a novel class of frozen initial conditions which prove capable of emulating a quantised primordial power spectrum.
\end{abstract}

\maketitle

\section{Introduction}\label{Introduction}
With only six parameters, the $\Lambda$CDM concordance model of the Universe explains the large-scale structure, present state, and evolution of the cosmos to high precision~\citep{LCDM}. Two of these parameters phenomenologically describe the amplitude $A_\mathrm{s}$ and tilt $n_\mathrm{s}$ of the primordial power spectrum of comoving curvature perturbations. The detection of $n_s\neq1$, along with correlated acoustic oscillations in the temperature and polarisation of the cosmic microwave background (CMB) anisotropies~\citep{planck_2013} constitutes overwhelming evidence for a rapid early accelerated phase. The canonical method for explaining this evolution is the theory of inflation, in which primordial quantum fields drive the accelerated expansion.

Models of inflation make predictions about the primordial power spectrum. These predictions may be tested against observations of the CMB, allowing us to probe the physics of this hypothesised embryonic stage of the universe~\citep{planck_inflation,planck_parameters}. Traditional analyses manifest these predictions in terms of $A_\mathrm{s}$ and $n_\mathrm{s}$ conditioned on inflationary model parameters. In many cases, such a brutal phenomenological parameterisation is insufficient. These cases include models that explain large scale features in CMB power spectra~\citep{WMAP1,WMAP2,WMAP3,Features}, axion monodromy models~\citep{monodromy}, just enough inflation~\citep{just_enough_inflation} or kinetic dominance~\citep{Kinetic,Hergt1,Hergt2}. Occasionally one may have access to analytic expressions or approximations, but in general one must solve the Mukhanov-Sasaki mode equations numerically in order to compute primordial power spectra. 

In many instances of cosmological interest the numerical computation of primordial power spectra forms the primary bottleneck in a full numerical inference.  There have been several attempts at tackling this problem \citep{BINGO, Galvez1, Galvez2} using a variety of analytical and numerical approaches with varying degrees of generality and efficiency. In this paper we present a novel and general approach to solving the Mukhanov-Sasaki equation. In all cases this approach gives a wavenumber-dependent increase in speed, and for large wavenumbers this method presents the only currently available method for computing fully numerical solutions that integrate across the entire evolution of the mode equations.

The format of this paper is as follows: In \cref{sec:background} we summarise relevant background theory and establish notation. In \cref{sec:methodology}, we describe the general approach. In \cref{sec:results} we apply our technique to the challenging case of a stepped potential with kinetic dominance initial conditions and compare with traditional solving techniques. We conclude in \cref{sec:conclusion}. 

\section{Background}\label{sec:background}

The simplest inflationary model is provided by a single scalar field minimally coupled to gravity with action
\begin{equation}
    S=\frac{1}{2}\int d^4x\sqrt{|g|}\big[R+\nabla^{\nu}\phi\nabla_{\nu}\phi - V(\phi)\big],
    \label{eq:action}
\end{equation}
where $\phi$ is a scalar inflaton field, $V(\phi)$ its potential, and we are working in natural units. For a flat Friedmann--Lema\^{i}tre--Robertson--Walker universe filled with a spatially uniform field $\phi$, extremising this action recovers the Klein-Gordon and Friedmann equations
\begin{gather}
    \ddot{\phi}+3H\dot{\phi}+V'(\phi)=0,
    \label{eqn:klein_gordon}\\
    H^2 = \frac{1}{3} \left(\frac{1}{2}\dot{\phi}^2 + V(\phi)\right),
    \label{eqn:friedmann}
\end{gather}
where $H=\dot{a}/a$ is the Hubble parameter, and dots and primes denote derivatives with respect to cosmic time $t$ and the field $\phi$ respectively.

For nearly all potentials, the background solutions to~\cref{eqn:klein_gordon,eqn:friedmann} converge on the {\em slow-roll\/} attractor solution satisfying $\dot{\phi}^2\ll V(\phi)$. In this regime, $\phi$ and $H$ are approximately constant and the Universe undergoes approximately exponential expansion $a\sim \exp(H t)$, causing the comoving horizon ${(aH)}^{-1}$ to shrink. One may set self-consistent slow-roll initial conditions~\cite{Hiranya_potential_1,Hiranya_potential_2,Hiranya_potential_3} via the constraint:
\begin{equation}
    \dot{\phi} = -\frac{V'(\phi)}{3H}, \quad 3H^2 = V(\phi),
    \label{eqn:slow_roll}
\end{equation}
allowing a small amount of evolution to remove transient effects resulting from an initial small offset from the true attractor.

Alternatively, one may set initial conditions in a kinetically dominated phase. Whilst late-time inflationary evolution is characterised by a slow-moving inflaton, classically at early times the opposite, $\dot{\phi}^2\gg V(\phi)$, is generally true~\citep{Kinetic}. In this state, the comoving horizon grows until the inflaton is sufficiently slowed by the friction term in \cref{eqn:klein_gordon}. A brief transitional {\em fast-roll\/} period $\dot{\phi}^2\sim V(\phi)$ is reached before the field settles into the usual slow-roll phase. Kinetic dominance initial conditions may be set via:
\begin{align}
    H(t) =\frac{1}{3t},\quad
    \phi(t) =\phi_{\rm p} \pm \sqrt{\frac{2}{3}}\log t,\quad
    a(t) \propto t^\frac{1}{3},
\end{align}
where $\phi_\mathrm{p}$ is a constant of integration.

\begin{figure}
    \includegraphics{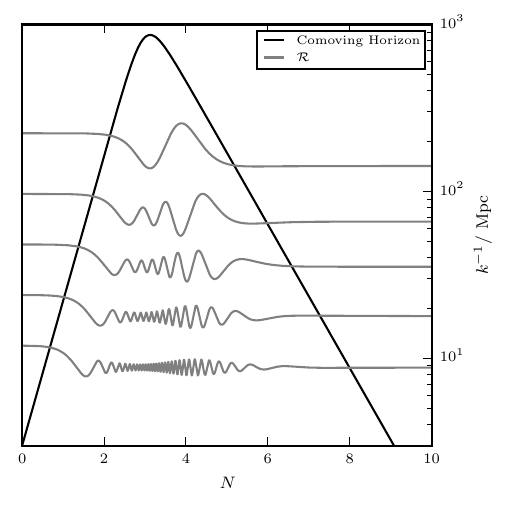}
    \caption{Comoving horizon along with appropriately shifted mode evolution $\mathcal{R}_k$ for different values of $k$. The comoving horizon expands in kinetic dominance until the beginning of inflation when it begins shrinking. The $k$-modes freeze-out as they exit the comoving horizon.}\label{fig:exit}
\end{figure}

Perturbing the action in \cref{eq:action} around the zeroth-order homogeneous solutions and taking scalar components yields the gauge-invariant Mukhanov action \cite{tasi} 
\begin{equation}
    S_{(2)}= \frac{1}{2}\int dt d^3x \big[\dot{\mathcal{R}}^2-a^{-2}(\partial_i \mathcal{R})^2\big]a z^2, \quad
    z = \frac{a \dot{\phi}}{H},
    \label{eqn:mukhanov_action}
\end{equation}
where $\mathcal{R}$ is the gauge-invariant comoving curvature perturbation. Varying this action and expressing $\mathcal{R}$ in terms of its isotropic Fourier components $\mathcal{R}_k$, we obtain the Mukhanov-Sasaki (MS) equation
\begin{equation} \label{eq:MS_R}
    \ddot{\mathcal{R}}_k + \left[H + 2\frac{\dot{z}}{z}\right]\dot{\mathcal{R}}_k + \frac{k^2}{a^2} \mathcal{R}_k = 0.
\end{equation}
As shown in \cref{fig:exit}, solutions to these equations are oscillatory within the horizon $k\gg aH$ and freeze out upon horizon exit. For tensor perturbations, the equivalent equation for both polarisations is
\begin{equation} \label{eq:MS_t}
    \ddot{\mathcal{T}}_k + 3H \dot{\mathcal{T}}_k + \frac{k^2}{a^2} \mathcal{T}_k = 0,
\end{equation}
In the slow-roll paradigm, initial conditions for the perturbations are typically set using Bunch--Davies initial conditions, by matching:
\begin{equation}
    \lim_{k\gg aH} \mathcal{R}_k=\frac{1}{z\sqrt{2k}}e^{-i k \int \frac{dt}{a}},\quad \mathcal{T}_k = \frac{2}{a\sqrt{2k}}e^{-i k \int \frac{dt}{a}}.
    \label{eqn:bunch_davies}
\end{equation}
In the fast-roll and kinetically dominated paradigms, this condition can never be fulfilled for small $k$. In these cases, the situation becomes less clear-cut, although alternative initial conditions have been proposed~\cite{VacuumChoices,Novel,Fruzsina}.

Once initial conditions have been chosen, to compute primordial power spectra one must evolve all modes of interest until they are well outside the horizon and evaluate
\begin{equation}
    \mathcal{P}_\mathcal{R}(k) = \lim_{k\ll aH}  {\frac{k^3}{2\pi^2}} |\mathcal{R}_k|^2,
    \quad\mathcal{P}_\mathcal{T}(k) = \lim_{k\ll aH}  {\frac{k^3}{2\pi^2}} |\mathcal{T}_k|^2.
    \nonumber
\end{equation}

For large values of $k$ many oscillations must be traversed in order to reach horizon exit, causing standard numerical solvers to fail. This may be somewhat ameliorated in the slow-roll case by starting the evolution a small amount before horizon exit, and exploit the fact that for slow-roll, Bunch-Davies conditions may be set anywhere within the horizon~\cite{ModeCode1,ModeCode2,ModeCode3}. However, such short-cuts can be harder to apply for certain potentials, particularly ones which yield spectra with moderately high-$k$ features.

In this paper, we choose the stepped quadratic potential~\cite{Features} as a challenging but relevant example
\begin{equation}
    V(\phi) = \frac{1}{2}m^2 \phi^2 \left[ 1 + A\tanh\left( \frac{\phi-\phi_0}{\Delta} \right) \right],
    \label{eq:potential}
\end{equation}
where $m$ is the mass of the inflaton field and $A$, $\phi_0$, and $\Delta$ are the amplitude, location and width of a step feature. Such step features induce oscillations in the primordial power spectrum,  which could be responsible for the low-$\ell$ features seen in CMB power spectra~\citep{WMAP1,WMAP2,WMAP3,Features}.

\section{Methodology}\label{sec:methodology}
We now review our new approach for evolving the Mukhanov-Sasaki mode \cref{eq:MS_R,eq:MS_t}, first in a general context in \cref{sec:transition_approach,sec:interval_choice}, and then in application to primordial cosmology in \cref{sec:MS_eq}.

\citet{Contaldi} construct an analytic template for the scalar primordial power spectrum with fast-roll initial conditions. They find exact solutions to \cref{eq:MS_R} in both the kinetic dominance and slow-roll limits and match them together assuming an instantaneous transition between kinetic dominance and slow-roll. This produces an expression in terms of Bessel functions which recovers the key features of the primordial power spectra with cut-offs and oscillations. In our approach, we increase the accuracy of the \citet{Contaldi} method by adding further transitions, allowing for an approximately continuous matching of fast roll to slow roll, and for the reconstruction of features within slow-roll inflation.

\subsection{The transition approach for an oscillator}\label{sec:transition_approach}
For a general linear second order differential equation of the form
\begin{equation}
    p(t)\ddot{y}(t) + 2q(t)\dot{y}(t) + r(t)y(t) = 0, 
    \label{eq:general_diff}
\end{equation}
one can find suitable dependent variable transformations that cast the differential into the form of a harmonic oscillator. Defining
\begin{equation}
    y(t) = x(t) \exp\left({-\int\frac{q(t)}{p(t)}dt}\right),
    \label{eq:transformation}
\end{equation}
\cref{eq:general_diff} may be cast as 
\begin{equation}
    \ddot{x}(t) + \omega^2(t)\: x(t) = 0,
\end{equation}
where
\begin{equation}
    \omega^2(t) =  \frac{r}{p} - \left(\frac{q}{p}\right)^2 - \frac{d}{dt}\left(\frac{q}{p}\right).
\end{equation}
A condition for the functionality of our method is that the integral $ \int q(t)/p(t) dt$ has an analytic expression or is numerically cheap to calculate and that $\omega^2(t)$ is a reasonably well-behaved function. There is also a freedom in choosing the independent variable which slightly modifies the form of \cref{eq:transformation} and $\omega^2(t)$. 

Thus, the Mukhanov-Sasaki equations may always be cast into the form of a harmonic oscillator. The form of $\omega^2(t)$ is {\em a priori\/} analytically unknown, and typically derived from inflationary background variables which are themselves numerical solutions of their own separate differential equation, as is demonstrated later in \cref{eq:omega_S,eq:omega_Q}. However, one may approximate the true frequency as a piecewise interpolation function. Since $\omega^2(t)$ may be negative and can span many decades in scale, we choose a semi-log interpolation function defined on $n$ intervals $\{[t_0,t_1),\ldots,[t_{n-1},t_n)\}$. For each interval one chooses either a linear, positive exponential or negative exponential parameterisation:
\begin{equation} \label{eq:w2}
    \omega^2(t) =
    \left\{
    \begin{array}{ll}
        a_i+b_i t & $Linear$  \\ 
        e^{2 (a_i + b_i t)} & +$Exp$ \\ 
        -e^{2 (a_i + b_i t)} & -$Exp$ \\ 
    \end{array}
    \right.
\end{equation}
where  
\begin{equation}
    b_i  = \frac{\omega^2_{i+1} - \omega^2_i}{t_{i+1}-t_i}, \quad
    a_i  = \omega^2_i - b_i t_i, \quad \omega^2_i = \omega^2(t_i)
\end{equation}
for the linear segments and
\begin{equation}
    b_i  = \frac{\ln|\omega^2_{i+1}|- \ln |\omega^2_{i}|}{2(t_{i+1}-t_i)}, \quad
    a_i  = \frac{\ln|\omega^2_i|}{2} - b_i t_i,
\end{equation}  
for the exponential segments. The choice of linear, positive or negative exponential segments is subject to the constraint that $\omega^2(t)$ must be purely positive or negative for the exponential regions.

The critical insight in this approach is that when $\omega^2(t)$ takes one of the three forms in \cref{eq:w2}, exact analytic solutions can be found in terms of Airy and Bessel functions
\begin{equation} \label{eq:Q_solutions}
    x(t) =  
    \left\{
    \begin{array}{llll}
        C_1 \operatorname{Ai}\left(-\frac{a_i+b_it}{|b_i|^{{2}/{3}}}\right) &+& C_2 \operatorname{Bi}\left(-\frac{a_i+b_it}{|b_i|^{{2}/{3}}}\right) & $Linear$ \\
        C_3 J_0\left(\frac{e^{a_i+b_it}}{|b_i|}\right) &+& C_4 Y_0\left(\frac{e^{a_i+b_it}}{|b_i|}\right)  & +$Exp$ \\
        C_5 I_0\left(\frac{e^{a_i+b_it}}{|b_i|}\right) &+& C_6 K_0\left(\frac{e^{a_i+b_it}}{|b_i|}\right)  & -$Exp$,
    \end{array}
    \right.
\end{equation}
where $C_i$ are constants of integration.

The full evolved solutions can be found by matching the value and first derivative of the solution at each transition boundary using matrix multiplication. First, define the following matrices 
\begin{align}
\mathcal{M}^{\sim}_{i} (t) &= 
\left[
    \begin{array}{cc}
        \mathcal{A}_i(t)&\mathcal{B}_i(t) \\ 
        \dot{\mathcal{A}_i}(t)&\dot{\mathcal{B}_i}(t) 
    \end{array} 
\right],
\nonumber\\ 
\mathcal{M}^{+}_{i} (t) &= 
\left[
    \begin{array}{cc}
        \mathcal{J}_i(t)&\mathcal{Y}_i(t) \\ 
        \dot{\mathcal{J}_i}(t)&\dot{\mathcal{Y}_i}(t) 
    \end{array} 
\right],
\nonumber\\
\mathcal{M}^{-}_{i} (t) &= 
\left[
    \begin{array}{cc}
        \mathcal{I}_i(t)&\mathcal{K}_i(t) \\ 
        \dot{\mathcal{I}_i}(t)&\dot{\mathcal{K}_i}(t) 
    \end{array} 
\right],
\end{align}
where the superscript $\sim, -, +$ indicates the transition type as linear, negative exponential, and positive exponential respectively, and
\begin{align} \label{eq:Def}
        \mathcal{A}_i(t) &= \operatorname{Ai}\left(-\frac{a_i + b_i t}{|b_i|^{{2}/{3}}}\right), 
        &\mathcal{B}_i(t) &= \operatorname{Bi}\left(-\frac{a_i + b_i t}{|b_i|^{{2}/{3}}}\right),  		\nonumber\\ 
        \mathcal{J}_i(t) &= J_0\left(\frac{e^{a_i + b_i t}}{|b_i|}\right), 
        &\mathcal{Y}_i(t) &= Y_0\left(\frac{e^{a_i + b_i t}}{|b_i|}\right), 
         \nonumber\\
        \mathcal{I}_i(t) &= I_0\left(\frac{e^{a_i + b_i t}}{|b_i|}\right),  
        &\mathcal{K}_i(t) &= K_0\left(\frac{e^{a_i + b_i t}}{|b_i|}\right).
\end{align}
The evolved solution from $t_0$ to $t_n$ can now be expressed in the compact form
\begin{align} \label{eq:Mat_CD}
\left[
    \begin{array}{r}
        x(t_n) \\ 
        \dot{x}(t_n) 
    \end{array} 
\right]
 &=   
\mathcal{U}(t_n,t_0) 
\left[
    \begin{array}{r}
x(t_0) \\ \dot{x}(t_0) 
    \end{array} 
\right],
\\
    \mathcal{U}(t_n,t_0)&=   
\prod_{i=0}^{n-1} \mathcal{M}^{j_i}_i(t_{i+1}) [\mathcal{M}^{j_i}_i(t_{i})]^{-1},
\end{align}
where the superscript $j_i \in \{\sim,+,-\}$ denotes the type of transition for the interval $[t_i,t_{i+1})$. The matrix $\mathcal{U}$ can be thought of as a linear evolution operator which acts on a state at time $t_0$ to evolve it to $t_n$.

\subsection{Interval choice}\label{sec:interval_choice}

The above argument in \cref{sec:transition_approach} was conditioned on a specific definition of intervals and interval types defining a semi-logarithmic interpolation of $\omega^2(t)$. In the limit of arbitrarily fine intervals, this approach recovers the exact solution. However, in order to minimise computational time, one should choose a coarser distribution of intervals with not necessarily constant width. In this section we outline one possible approach for making such a choice.

One may approximate a local error in the solution $x$ across each interval $[t_i, t_{i+1})$ by computing solutions at either end, and then repeating the calculation across two adjacent and matched intervals $[t_i,t_m),[t_m,t_{i+1})$, where $t_m = (t_i + t_{i+1})/2$ is the midpoint of the original interval. The difference in these two approaches gives a rough quantification of the local error accumulated from $t_i$ to $t_{i+1}$. If the error is greater than some user-specified tolerance, then the interval is bisected, and the above process is repeated on each of the two segments. For our application, $x$ is in general complex, and we quantify the local error as a relative error between absolute values of the two alternative solutions.

To choose initial segments which are then refined by the above procedure, we select $t_0,\ldots,t_n$ to be the endpoints of our region of interest, along with the locations of extrema of $\omega^2(t)$. Including extrema ensures that no sharp features are missed. The interpolation type for each transition is selected to give the lowest error in $x$.

\subsection{The Mukhanov-Sasaki equation}\label{sec:MS_eq}

\begin{figure} 
    \includegraphics{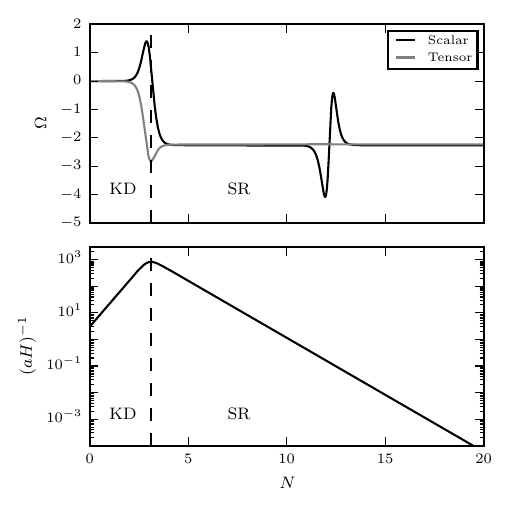}
    \caption{The two component terms of $\omega^2_k$ from \cref{eq:Q,eq:S,eq:omega_Q,eq:omega_S}. Both terms have two distinct regions. The first region is kinetically dominated. The second is the slow-roll region which is slowly varying. The feature in the slow-roll region is caused by the step in the potential from \cref{eq:potential}.}\label{fig:w2}
\end{figure}

In order to apply the method outlined in \cref{sec:transition_approach,sec:interval_choice} to the computation of the primordial power spectrum of comoving curvature perturbations, we must first recast the Mukhanov-Sasaki \cref{eq:MS_R,eq:MS_t} in a more appropriate form. As observed by \citet{Fruzsina}, there is considerable freedom in the form that these equations take, as one may simultaneously transform both the independent and dependent variables. 

As an independent variable, we choose the number of $e$-folds $N = \text{ln}\left(\frac{a}{a_0}\right)$. Provided $H>0$, $N$ constitutes a stable temporal coordinate that does not saturate during inflation, and naturally pushes the kinetic dominance singularity to $N=-\infty$. Requiring that there is no friction term, we are forced to choose the following transformations and equations:
\begin{align}
    \mathcal{S} = z\sqrt{aH} \mathcal{R}&\Rightarrow&
    \frac{d^2 \mathcal{S}_k}{dN^2} + \left[\Omega_\mathcal{S} + \frac{k^2}{a^2H^2}\right] \mathcal{S}_k = 0,
    \label{eq:Q}\\
    \mathcal{Q} = a\sqrt{aH} \mathcal{T} &\Rightarrow& 
    \frac{d^2 \mathcal{Q}_k}{dN^2} + \left[\Omega_\mathcal{Q} + \frac{k^2}{a^2H^2}\right] \mathcal{Q}_k = 0,
    \label{eq:S}
\end{align}
where $\omega^2_k$ is separated into a $k$-dependent part, which is proportional to the square of the comoving horizon, and a $k$-independent part:
\begin{align}
    \Omega_\mathcal{S} &= 
    \frac{V''}{H^2} + \frac{3}{2}  \frac{V' }{H^2} \frac{d\phi}{dN} - \frac{1}{16} \left(\left(\frac{d\phi}{dN}\right)^2 - 6\right)  \left(5 \left(\frac{d\phi}{dN}\right)^2 - 6\right),
    \label{eq:omega_Q}\\
    \Omega_\mathcal{Q} &= 
    -\frac{1}{2} \frac{V'}{H^2}\frac{d\phi}{dN}  + \frac{1}{16}  \left( \left(\frac{d\phi}{dN}\right)^2 - 6\right) \left(3\left(\frac{d\phi}{dN}\right)^2 + 6\right),
    \label{eq:omega_S}
\end{align}
illustrated graphically in \cref{fig:w2}. 

\subsection{Comparison with existing approaches}

Traditional solvers such as \texttt{ModeCode}~\cite{ModeCode1,ModeCode2,ModeCode3} and \texttt{BINGO} \cite{BINGO} are able to avoid computing a full numerical evolution by starting the mode evolution for each mode shortly before horizon exit. This proves sufficient for many physical situations, since deep within the horizon $k\gg aH$ many traditional initial conditions reduce to the Bunch-Davies vacuum, allowing careful analyses to skip the large number oscillations required to reach horizon exit. In many ways, our approach can be thought of an automation of this skipping procedure via its switching mechanism. Furthermore, the approach outlined here allows one to investigate a wider variety of initial conditions, such as excited states or alpha-vacua~\cite{alpha_vacua,2012JCAP...12..012C, 2012JCAP...04..039D, 2011JCAP...03..025A, 2005JCAP...08..001E, 2005JHEP...05..063B, 2005tsra.conf....1G, 2005PhRvD..71b3516D, 2004PhRvD..70h4019C, 2003PhRvD..68l4012C, 2003NuPhB.669..325G, 2002JHEP...11..037K, 2002PhRvD..66b3511D}, which require full evolution of mode functions from early times. 

\section{Results}\label{sec:results}

A C++ implementation of the approach outlined in \cref{sec:methodology} is publicly available on \texttt{GitHub}~\cite{GitHub}. We make use of pre-packaged libraries \texttt{ODEPACK}~\cite{odepack} for solving ordinary differential equations, \texttt{cephes}~\cite{cephes} for special functions and \texttt{Eigen}~\cite{eigen} for vector arithmetic. We compare our approach in speed and accuracy to a full numerical solution of the Mukhanov-Sasaki equation using \texttt{ODEPACK}. This numerical approach is analagous to \texttt{ModeCode}~\cite{ModeCode1,ModeCode2,ModeCode3} and \texttt{BINGO} \cite{BINGO}, whereby mode evolution is started and stopped a short way before and after horizon exit to minimise run-time. 

\subsection{Kinetic initial conditions}

\begin{figure}
    \includegraphics{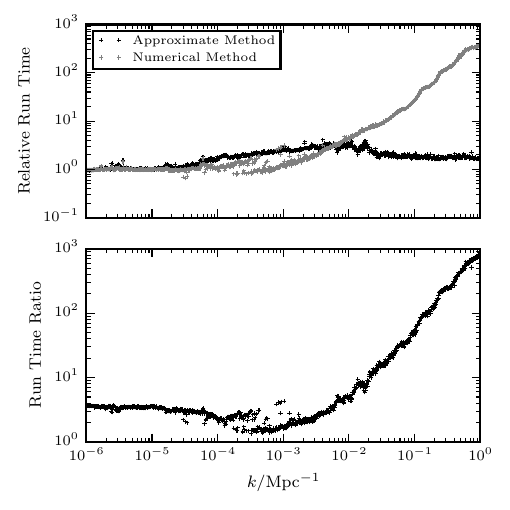}
    \caption{{\em Upper panel:\/} time taken per $k$-mode relative to $k=10^{-6} \mathrm{Mpc}^{-1}$ using the transition-based method and the traditional approach. {\em Lower panel:\/} speed ratio of the traditional approach relative to the transition-based method. The transition based  method is always faster per-$k$ than the traditional approach, and is orders of magnitude faster for large $k$.}\label{fig:speed}
\end{figure}

\begin{figure}
    \includegraphics{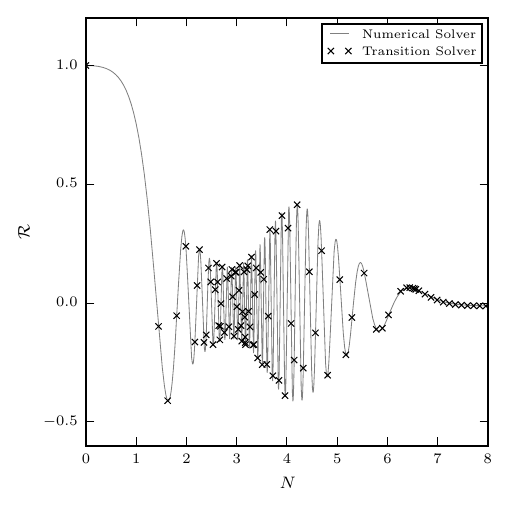}
    \caption{
        Example mode function evolution, transition points chosen for a tolerance of $10^{-3}$. Note that in fact, our solver uses $\mathcal{S}$, which is related to $\mathcal{R}$ via \cref{eq:Q}, and we only plot the real part for $k = 0.005$, $N_* = 55$, $N_\dagger = 7$.
    }\label{fig:R_Evolution}
\end{figure}

\begin{figure}
    \includegraphics{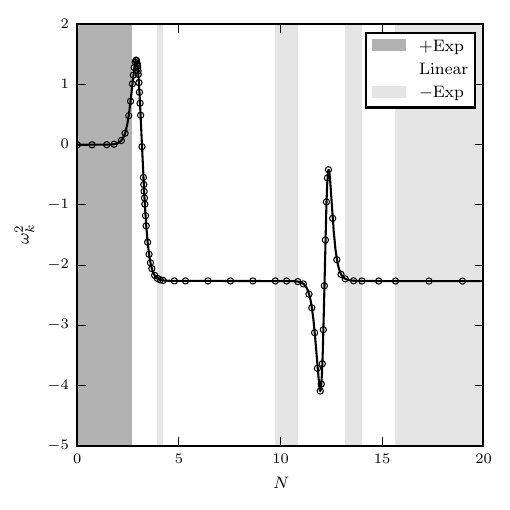}
    \caption{The exact solution of $\omega^2_k$ for scalar perturbations at $k=10^{-6}$ Mpc$^{-1}$ plotted with multiple transition approximations. The transitions have been selected using the method described in \cref{sec:interval_choice} and a fixed fractional error tolerance of $10^{-2}$.}				\label{fig:Transitions}
\end{figure}

To demonstrate the robustness of our approach, we apply it to the evolution resulting from the stepped potential from \cref{eq:potential}. The relative speed increase can be seen in \cref{fig:speed}. At high $k$ there are orders of magnitude increase in speed, and the method is approximately constant in cost for each $k$, in stark contrast to the traditional approach. An example mode evolution along with the transitions the solver chooses can be seen in \cref{fig:R_Evolution,fig:Transitions}. Our solver is able to navigate the oscillatory regions more effectively than an equivalent traditional Runge-Kutta based approach, as used in \texttt{ODEPACK}, and this effectiveness increases with $k$. 

\begin{figure}
    \includegraphics{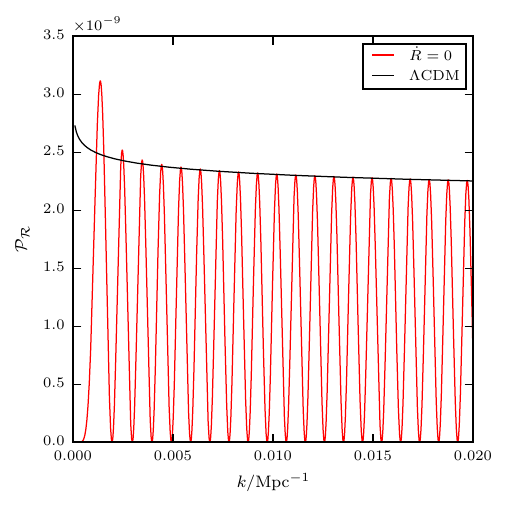}
    \includegraphics{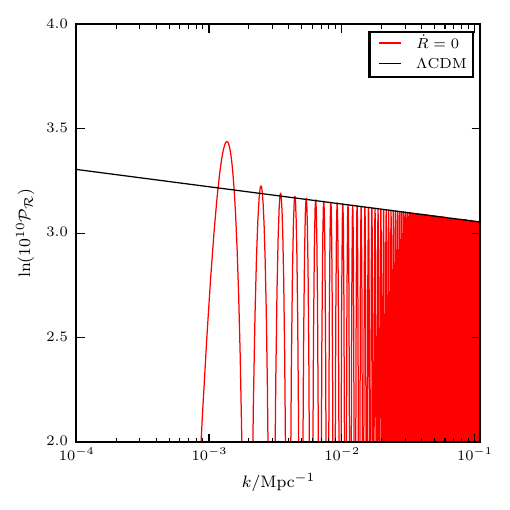}
    \caption{
        $\phi^2$ potential with $m = 2.9e-6$.
        $N_* = 55$
        $N_\dagger = 4.7$
        $N_r$  set as far back as possible.
    }\label{fig:quantised_pps}
\end{figure}

\begin{figure*}
    \includegraphics{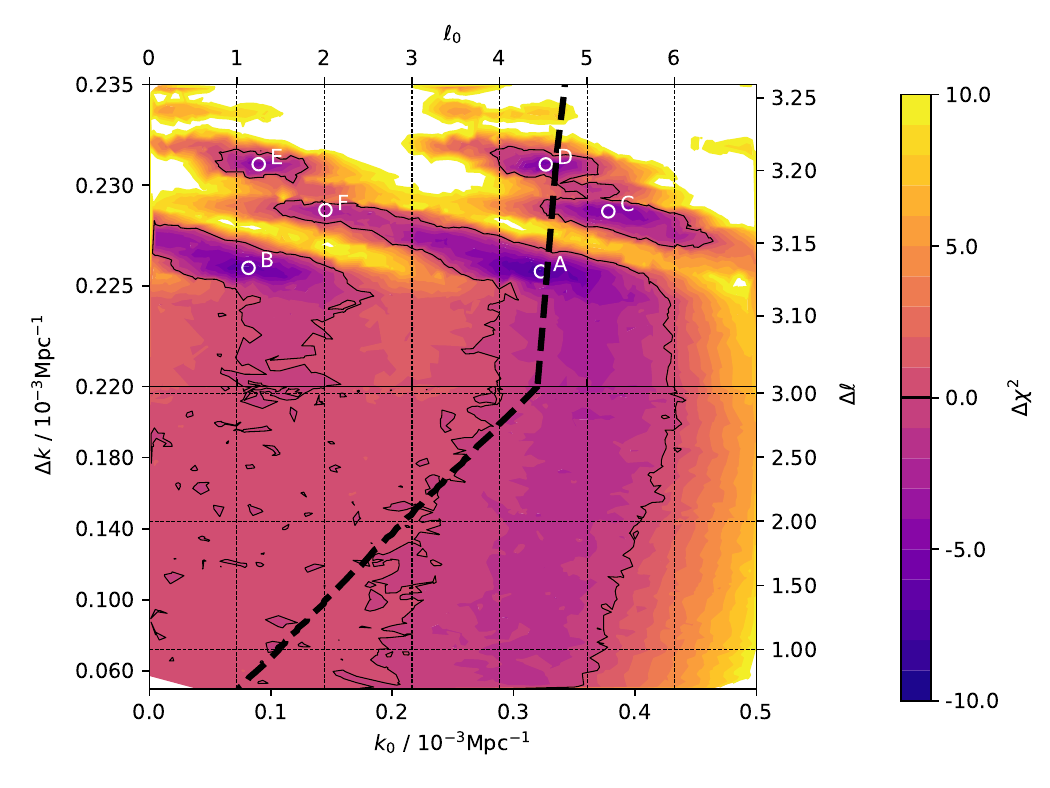}
    \caption{Plot reproduced from~\cite{bartlett} showing the quality of fit of a general quantised power spectrum with linear spacing $\Delta k$ and starting wavevector $k_0$. The corresponding approximate multipole spacing $\Delta \ell$ and initial multipole $\ell_0$ are also indicated. Frozen initial conditions predict a class of allowed $(k_0, \Delta k)$, shown by the dashed line, which independently predicts the best fit point in this class.
    }\label{fig:contour}
\end{figure*}

\begin{figure*}
    \includegraphics{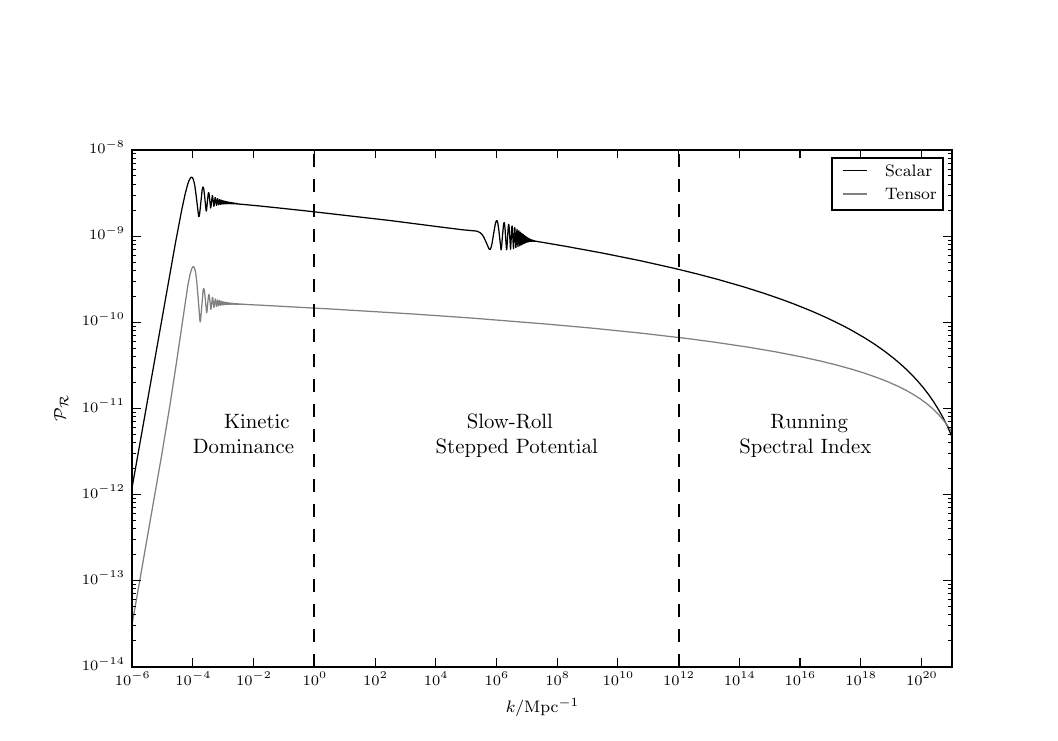}
    \caption{Power spectra of scalar and tensor perturbations computed using the method described in \cref{sec:methodology}. The background variables were computed using a stepped potential ($A=10^{-3}$, $\Delta=5\times10^{-3}$, $\phi_0 = 12.5$) and under kinetic dominance initial conditions with $N_\dagger = 7$ and $N_* = 55$. The spectra were computed under Bunch-Davies initial conditions and the vacuum was set $N_0 = 63.1$ $e$-folds before the end of inflation. The plot shows three different regions of the two spectra: A kinetic dominance initial region, slow-roll with a stepped potential targeting $k = 10^{7} $Mpc$ ^{-1}$, and a high $k$ region illustrating the running of the spectral index. As seen from \cref{fig:w2}, the feature in $\omega^2_k$ for the tensor perturbations is much smaller than that of the scalar perturbations. This is reflected in the magnitude difference of the oscillations caused by the step. The fractional error was below $0.1\%$ in both spectra.}\label{fig:power}
\end{figure*}

Using the notation of \citet{Hergt2}, power spectra are phenomenologically characterized by three parameters: 
\begin{description}
    \item[$N_*$] $e$-folds between the moment the pivot scale $k_*$ exits the horizon and the end of inflation,
    \item[$N_\dagger$] $e$-folds between the start of inflation and the horizon exit of pivot scale,
    \item[$N_0$] $e$-folds between the time that initial conditions are set and the end of inflation.
\end{description}
We initialise the perturbation variables in the mode \cref{eq:MS_R,eq:MS_t} using Bunch-Davies vacuum initial conditions, and for demonstration purposes choose $N_\dagger=7$, $N_*=55$ and $N_0=63.1$. The resultant power spectra can be seen in \cref{fig:power}, and are divided into three regions: Kinetic dominance, stepped feature and a running spectral index.

The cut-off and oscillatory behaviour caused by kinetic dominance can be seen for low $k$-modes. The middle region shows the spectrum at moderately high $k$ after exiting kinetic dominance and settling into slow-roll. An oscillatory feature can be seen at $k\sim 10^{7} \mathrm{Mpc}^{-1}$ caused by the step in the potential of \cref{eq:potential}.
At high $k$, running of the spectral index $n_s$ can be seen as the spectrum tilts downwards. Whilst these $k$-modes affect multipoles that are too large to be probed directly by the CMB, they are relevant for example for the study of primordial black holes~\cite{highk}. 

\subsection{Frozen initial conditions}

To demonstrate the power of our approach, we now apply the method to a novel class of ``frozen initial conditions''. In some sense it would be attractive to remove the parameter $N_0$ from our initial conditions, and set $N_0\to-\infty$ with initial conditions asymptotically deep within the kinetically dominated phase ``at the big bang''. In this case, to avoid modes growing backward in time (so that the perturbative approximation remains valid~\cite{anthony}), one must select the initial conditions:
\begin{equation}
    {|\mathcal{R}_k|}^2 \propto 1,\qquad \dot{\mathcal{R}}_k = 0.
    \label{eqn:frozen}
\end{equation}
These initial conditions are illustrated graphically in \cref{fig:exit}, and amount to a white noise pre-primordial power spectrum. A direct consequence of these initial conditions is that the modes are purely real, and therefore exhibit an acoustic oscillation-like effect upon horizon re-entry. The resulting power spectrum is visualised in \cref{fig:quantised_pps}, which shows heavy oscillations down to zero power. These oscillations in the primordial power spectrum are akin to the quantised primordial power spectra that have recently been examined in~\cite{anthony,bartlett}, however in this case our initial conditions predict that both the smallest wavevector $k_0$ and quantisation spacing $\Delta k$ are functions of $N_\dagger$. Compellingly, \cref{fig:contour} shows that this class of $(k_0,\Delta k)$ comprise a curve that slides directly through the best-fit point found in \cite{bartlett}. This model therefore provides the possibility of a significantly improved fit in comparison to $\Lambda$CDM with the introduction of a only a single additional parameter.

These primordial power spectra can only be computed numerically by a solver such as ours which is capable of navigating the many oscillations between horizon entry and exit. Given the completely independent prediction of this best fit point by these initial conditions, ``Frozen initial conditions'' will form the subject of a future paper which examines the theoretical and full observational implications.

\section{Conclusion}\label{sec:conclusion}

In this paper, we described a novel method for the numerical calculation of the primordial power spectra of comoving curvature perturbations. The results were shown to agree well with existing numerical solutions of the Mukhanov-Sasaki equation ($0.1\%$ errors) while only requiring a fraction of the computational time.

With this fast and efficient method for calculating power spectra, further investigations into vacuum initial conditions can be explored and their effects on CMB power spectra can be tested and compared with observations. We plan to incorporate the code presented in~\cite{GitHub} as a CLASS extension~\cite{class}. 

Our approach is analogous to the Runge-Kutta-Wentzel-Kramers-Brillouin method~\cite{RKWKB}, differing in its choice of stepping function and error control. As for \mbox{RKWKB}, there is much scope for extensions to our method, including but not limited to higher-order stepping procedures and the integration of coupled oscillators. On the inflationary physics side there is also scope to extend this work to multi-field inflation, non-minimally coupled inflation, and spatial curvature.

\begin{acknowledgements}
    WIJH was supported by a King's College studentship and by the Kavli foundation. WJH was supported by a Gonville \& Caius research fellowship.
\end{acknowledgements}

\bibliographystyle{unsrtnat}
\bibliography{references}

\end{document}